\documentclass[10pt,conference]{IEEEtran}
\IEEEoverridecommandlockouts
\usepackage{amsmath,amssymb,amsfonts}
\usepackage{tcolorbox}
\tcbuselibrary{listings, breakable}
\usepackage{adjustbox}
\usepackage{enumitem}
\usepackage{graphicx}
\usepackage{textcomp}
\usepackage{xcolor}
\usepackage{amsmath, amssymb}
\usepackage{mathtools}
\usepackage{algorithm}
\usepackage[inkscapelatex=false]{svg}
\usepackage{hyperref}

\usepackage{capt-of}
\newcommand{\mycomment}[1]{}
\usepackage{float}

\DeclarePairedDelimiter\ket{\lvert}{\rangle}
\DeclarePairedDelimiterX\braket[2]{\langle}{\rangle}{#1\,\delimsize\vert\,\mathopen{}#2}
\def\BibTeX{{\rm B\kern-.05em{\sc i\kern-.025em b}\kern-.08em
    T\kern-.1667em\lower.7ex\hbox{E}\kern-.125emX}}
\begin{document}

\title{Quantum-Enhanced Weight Optimization for Neural Networks Using Grover’s Algorithm\\
}

\author{\IEEEauthorblockN{\c Stefan-Alexandru Jura}
\IEEEauthorblockA{\textit{Department of Computers and Information Technology} \\
\textit{Politehnica University Timi\c{s}oara}\\
Timi\c soara, Romania \\
stefan.jura@student.upt.ro}
\and
\IEEEauthorblockN{Mihai Udrescu}
\IEEEauthorblockA{\textit{Department of Computers and Information Technology} \\
\textit{Politehnica University Timi\c{s}oara}\\
Timi\c soara, Romania \\
mihai.udrescu@cs.upt.ro}

}

\maketitle

\begin{abstract}
The main approach to hybrid quantum-classical neural networks (QNN) is employing quantum computing to build a neural network (NN) that has quantum features, which is then optimized classically. Here, we propose a different strategy: to use quantum computing in order to optimize the weights of a classical NN. As such, we design an instance of Grover's quantum search algorithm to accelerate the search for the optimal parameters of an NN during the training process, a task traditionally performed using the backpropagation algorithm with the gradient descent method. Indeed, gradient descent has issues such as exploding gradient, vanishing gradient, or convexity problem. Other methods tried to address such issues with strategies like genetic searches, but they carry additional problems like convergence consistency. Our original method avoids these issues—because it does not calculate gradients—and capitalizes on classical architectures' robustness and Grover's quadratic speedup in high-dimensional search spaces to significantly reduce test loss (58.75\%) and improve test accuracy (35.25\%), compared to classical NN weight optimization, on small datasets. Unlike most QNNs that are trained on small datasets only, our method is also scalable, as it allows the optimization of deep networks;  for an NN with 3 hidden layers, trained on the \emph{Digits} dataset from scikit-learn, we obtained a mean accuracy of 97.7\%. Moreover, our method requires a much smaller number of qubits compared to other QNN approaches, making it very practical for near-future quantum computers that will still deliver a limited number of logical qubits.
\end{abstract}


\section{Introduction}
\label{sec:intro}
Since the dawn of Deep Learning (DL) in 2012, one of its most investigated topics has been the convergence of NN \cite{b3}. Traditionally, we may achieve convergence after minimizing the loss function by calculating the loss's derivative for the network's initial weights. This way, the gradient descent method determines the optimal parameter values \cite{b4}. There are several variants of gradient descent in the literature: stochastic gradient descent, batch gradient descent, and mini-batch gradient descent \cite{b2, b5}. The literature also suggests variations of these methods that perform better than the original ones \cite{b2} in terms of convergence and reduced computational cost.

The convergence problem occurs because the loss function can be nonconvex; therefore, training the model to get stuck to a local minimum \cite{b7}. ADAM (\textit{Adaptive Moment Estimation}) is a gradient descent method that computes the first-order gradient (a vector that contains the first-order derivatives of the loss function for the model weights) and updates the weights using an adaptive learning rate \cite{b2}; when the loss function $f:\mathbb{R}^n \rightarrow \mathbb{R}$ is not convex, i.e., $\forall x,y \in \mathbb{R}^n$ and $\theta \in [0,1]$, $f\left(\theta x+(1-\theta)y\right) \leq \theta f(x)+(1-\theta)f(y)$, ADAM may fail to reach global optima, often converging to local optima and becoming trapped there \cite{b8}. The enhanced version of ADAM, BGADAM (\textit{Boosting Genetic ADAM}) \cite{b6}, addresses the convexity problem by combining the advantages of ADAM and metaheuristic genetic algorithms, helping the model escape from local optimal points and improving overall optimization. However, despite being efficient in finding reasonable solutions, genetic algorithms are not guaranteed to converge to global optima. First, poor population diversity can lead to premature convergence; second, the search space is partially sampled due to the small population; and third, genetic operators have a degree of randomness \cite{b8,b6,b16}.

We propose a method that uses quantum computing to address weight optimization in NN. To this end, we utilize Grover's algorithm, which brings a quadratic speedup (complexity $\mathcal{O}(\sqrt{N})$) over the classical unstructured space searches \cite{b9}; this is critical when working with large data sets and improves the learning capacity of a NN. Our method assigns weights the values of a uniform distribution---the spread of which should be as wide as possible because of the uncertainties in the optimal weight ranges. Then, for every weight, we define the search region around its present value, consisting of a finite number of sample values of the same distribution. We subsequently select the weight values that minimize the loss function using a modified version of Grover's algorithm that identifies the minimum value in an unstructured search space \cite{b10}. When our method finds the minimum loss, it updates the current weights to the newfound values, and then the search space is narrowed or expanded, depending on the value of the loss function. This iterative process converges to better weight values while avoiding getting stuck at a local optimum.

We developed and simulated our algorithm in Python using the NumPy library for numerical operations and the Qiskit framework for quantum circuit design and simulations. Since simulated quantum circuits and algorithms are computationally intensive, we performed simulation on a small number of qubits and a low search space resolution. This way, we made all proposed experiments computationally feasible while demonstrating the algorithm's potential for scarce computational resources. Larger-scale applications require larger quantum simulators or actual quantum hardware. 

Accordingly, the main contributions of this paper are:
\begin{itemize}

\item We introduce a new gradient-free optimization method in neural networks based on Grover's algorithm;

\item Compared to existing QNN approaches, our quantum optimization reduces the average test loss by 58.75\% and improves the average test accuracy by 35.25\% on small datasets; 

\item Owing to the scalability of our strategy, unlike other QNNs, we can implement deep networks, achieving better results than the classical multilayer perceptron.
\end{itemize}

This reminder of this paper is structured as follows: Section \ref{sec:background} reviews the related work, Section \ref{sec:arh} details the implementation of our NN optimization, Section \ref{sec:exp} compares the performance of our network with other optimization approaches, and Section \ref{sec:concl} provides concluding remarks.

\section{Background and Related Work}
\label{sec:background}
Our paper addresses weight optimization in NNs using a quantum algorithm. Therefore, the first subsection briefly introduces the basic concepts of DL and presents the algorithmic framework we used for optimization; the second subsection summarizes the state-of-the-art in quantum neural networks.
\subsection{Deep Learning and Quantum Computing} 
DL models have developed numerous applications over the years. For example, MLPs (\textit{Multi-Layer Perceptrons}) are designed for regression and classification problems \cite{b11}; they consist of several layers linked using activation functions, such as ReLU, hyperbolic tangent, or sigmoid \cite{b12,b13}; this helps the network to learn more complex patterns, especially when the data exhibit non-linear dependencies. After performing a forward pass through the network, the softmax function facilitates the interpretation of NN's outputs as probabilities that an input belongs to a specific class \cite{b27}.

Grover's search provides a quadratic speed-up of $\mathcal{O}(\sqrt{N})$ over classical methods \cite{b9,b23}; it uses an oracle $O$ to mark states representing search solutions by flipping their amplitude signs \cite{b22,b25,b26}; then, an iterative amplitude amplification increases the probability of measuring the correct result \cite{b23,b25,b26,b20}. Grover's algorithm can also be used to find the minimum or maximum in an unstructured state space \cite{b10}, i.e., a pool $P$ \cite{b1}. To identify the minimum---a crucial step in optimizing the network's loss function---the oracle $O$ is a function that marks all basis states with values lower than the provisional minimum $P_j$. Then, by amplifying the amplitudes of values lower than $P_j$ and quantum measurement, $P_k$ ($P_k< P_j$) becomes the new provisional minimum \cite{b22,b24,b30}. This procedure is repeated until it cannot render a value lower than the previous provisional minimum. 

\mycomment{\vspace{0.2cm}
\noindent\textbf{Quantum Algorithm for Finding the Minimum from an Unsorted Table of $m$ Elements}

\noindent 1. Define the oracle.
\begin{equation}
    O_j(i)=\begin{cases}
        1,\text{ if } P_i<P_j;\\
        0,\text{ otherwise.}
    \end{cases}
\end{equation}
\noindent2. Initialize $k:=\textit{random number}$, where $k \in \overline{0,m-1}$, as the starting point for this search. 

\noindent 3. Iterate \( \mathcal{O}(\sqrt{m}) \) times:
\begin{enumerate}
    \item[\hspace{1cm} a.] Set two quantum registers as an index-value pair, i.e., $\displaystyle{\ket{\psi}=\frac{1}{\sqrt{m}}\sum_{i=0}^{m-1} \ket{i}\ket{k}}$. The register $\ket{i}$ is a superposition of all the indices and the register $\ket{k}$ is the superposition of all the values.
    \item[\hspace{1cm} b.] Use Grover's algorithm to find marked states in the first register (that is, those that satisfy $O_k(i)=1$).
    \item[\hspace{1cm} c.] Measure the first register. The outcome will be one of the basis states that are indices for values less than $P_k$. Let the measurement result be x. Thus, make $k:=x$.
\end{enumerate}
\noindent 4. Return $k$ as the result; it is the index of the minimum.}

\subsection{Quantum Neural Networks}

Variational quantum circuits (VQCs), also referred to as variational quantum algorithms (VQAs), are a type of parametrized quantum circuits whose parameters are fine-tuned with a classical method (e.g., gradient descent) \cite{b31}. These quantum devices execute the circuit for some predefined parameters, measure the qubit value and return it to a classical computer where gradient descent optimization occurs \cite{b32}. Some of the applications of this circuit are the variational quantum eigensolver VQE (which estimates the ground-state energy of a physical system's Hamiltonian \cite{b33}), the approximate quantum optimization algorithm (tackles combinatorial optimization, e.g., the maximum cut algorithm \cite{b34}), and quantum machine learning (uses VQCs as quantum neural networks for classification or regression purposes \cite{b31}).

The existing literature on quantum neural networks often describes hybrid systems in which a variational quantum algorithm (VQA) mimics the structure of classical neural networks, with its parameters updated using classical optimization methods to minimize a cost function \cite{b35}. In such settings, Grover's algorithm can speed up convergence \cite{b35}. In contrast to classical perceptrons---where an activation function is applied to the sum of weighted inputs---in the quantum version, the input data are first encoded into a quantum state. A trainable quantum circuit (variational ansatz) then processes this state, and a subsequent qubit measurement produces probabilities that correspond to class predictions \cite{b35,b36,b38}. Quantum variational perceptrons (QVPs) have achieved superior classification accuracy in small datasets, underscoring the potential of quantum variational methods to enhance machine learning, particularly in scenarios where data are scarce and classical methods risk overfitting or failing to capture complex decision boundaries \cite{b36,b37}.

The variational (trainable) circuit $U(\theta)$ consists of CNOT gates that make the entanglement process possible and single-qubit rotations, such as $R_y(\theta)$ or $R_z(\theta)$, where $\theta$ is the angle of rotation around a specific axis on the Bloch sphere. After preparing \(\lvert \psi(x)\rangle\) (transforming the classical input $x$ into a quantum state), U($\theta$) is applied. For binary classification problems, we often measure one output qubit, as it can encode two quantum states, precisely the number of classes we have in our classification problem (class 0 and class 1) \cite{b36}. 
The classical gradient descent method optimizes the rotation angle $\theta$. Using cross-entropy loss $L$ for classification \cite{b15}, we update the angle using the learning rate $\eta$ and the gradient descent update formula.
In an accelerated variant, instead of using Grover's algorithm for unstructured space search, the authors add the Grover amplification stage after the variational circuit so that the overall quantum circuit is modified to boost the amplitudes of the quantum states corresponding to correct classifications, achieving faster convergence and higher accuracy \cite{b36}.

Another architecture represents a neuron by a single qubit, which evolves and interacts with the noisy environment \cite{b39}. Each quantum neuron of the previous layer is measured, yielding a 0 or 1; then, the neuron $j$ from the current layer receives each neuron $i$ output from the previous state. If the output of $i$ is $1$, a quantum operator $W_{ij}$ is applied to the qubit $j$; if the output of $i$ is 0, no operation is performed. The quantum operator $W_{ij}$ is a unitary gate, e.g., a rotation $R(\theta, \psi)$, which can be updated via gradient descent to minimize the loss function, similarly to QVP or the classical approach \cite{b39}.

The Online Quantum Perceptron (OQP) speeds up training by combining classical processing with quantum search. In OQP, each input is handled clasically by computing the dot product of the input with the weight vector and applying the sign function to predict +1 or –1. The quantum edge comes during error detection: rather than examining every training example in sequence (a $\mathcal{O}(N)$ task), OQP uses Grover’s search to locate a misclassified example in $\mathcal{O}(\sqrt{N})$ steps. It does so by preparing a superposition over training indices, using an oracle to flip the phase of indices corresponding to errors, and then amplifying those states with Grover diffusion. When an error is found, the perceptron updates its weights using $w \leftarrow w + y\cdot\theta$. This process repeats until all examples are correctly classified or a stopping criterion is met \cite{b28}.

The Quantum Version Space Perceptron (QVSP) adapts the classical support vector machines (SVM) by encoding the \emph{version space}---the set of hyperplanes that correctly classify the data---into a quantum superposition. Classically, finding a hyperplane with a margin $\gamma$ requires checking on the order of $1/\gamma^2$ candidates, because a small margin forces a finer search. In QVSP, an oracle marks the hyperplanes that classify correctly, and Grover's diffuser amplifies these marked states so that after $\mathcal{O}(1/\sqrt{\gamma})$ iterations, a measurement will likely yield a valid hyperplane. This brings a quadratic speedup over classical SVM \cite{b28}.

The QNN methods we reviewed here use Grover's algorithm to achieve speedup, but they do not employ Grover's algorithm for gradient-free weight optimization. Unlike QQP and QVSP, which update the weights' values based on the outputs without considering the loss function, our approach maintains the classical concept of an MLP while using Grover's algorithm to minimize the loss function. Moreover, we rely on an iterative update mechanism, which allows us to optimize an objective function rather than just correcting weights based on misclassified samples. Our NN architecture is classical but quantum-optimized, contrasting with most QNNs made of quantum gates and operators and optimized with classical methods. In a classical MLP, backpropagation is costly because derivatives are computed via a computational graph, which requires a lot of memory and can cause problems such as an exploding gradient or a vanishing gradient \cite{b15}. Our method circumvents the convexity problem because it does not calculate gradients.

\section{Proposed network architecture}
\label{sec:arh}

 We use an MLP developed for a classification task. Due to restrictions on power, the size of the
classical hardware, and the exponential memory complexity
needed to implement Grover's quantum algorithm, the simulation in Google Colaboratory is computationally prohibitive; this is because
classical hardware uses two classical bits to encode each qubit,
 meaning that the memory usage increases exponentially with the number of qubits.
Therefore, our network is purposely kept small: one
input layer, one hidden layer with 32 nodes (units or neurons), and one output layer. During training, we used regularization and dropout of $L_2$ ($p = 0.2$) to avoid numerical instability and overfitting. These
techniques enable the optimization process to focus on patterns
in the data that are more likely to be generalizable to new instances. However, as mentioned in Section \ref{sec:intro}, on a more powerful machine, we could implement a deep NN containing 3 hidden layers with 64, 32, and 16 units, respectively, showing the scalability and potential integration of our algorithm in actual AI applications.

\begin{figure*}[h]
    \centering
    \includegraphics[width=0.75\textwidth,height=0.5\textheight]{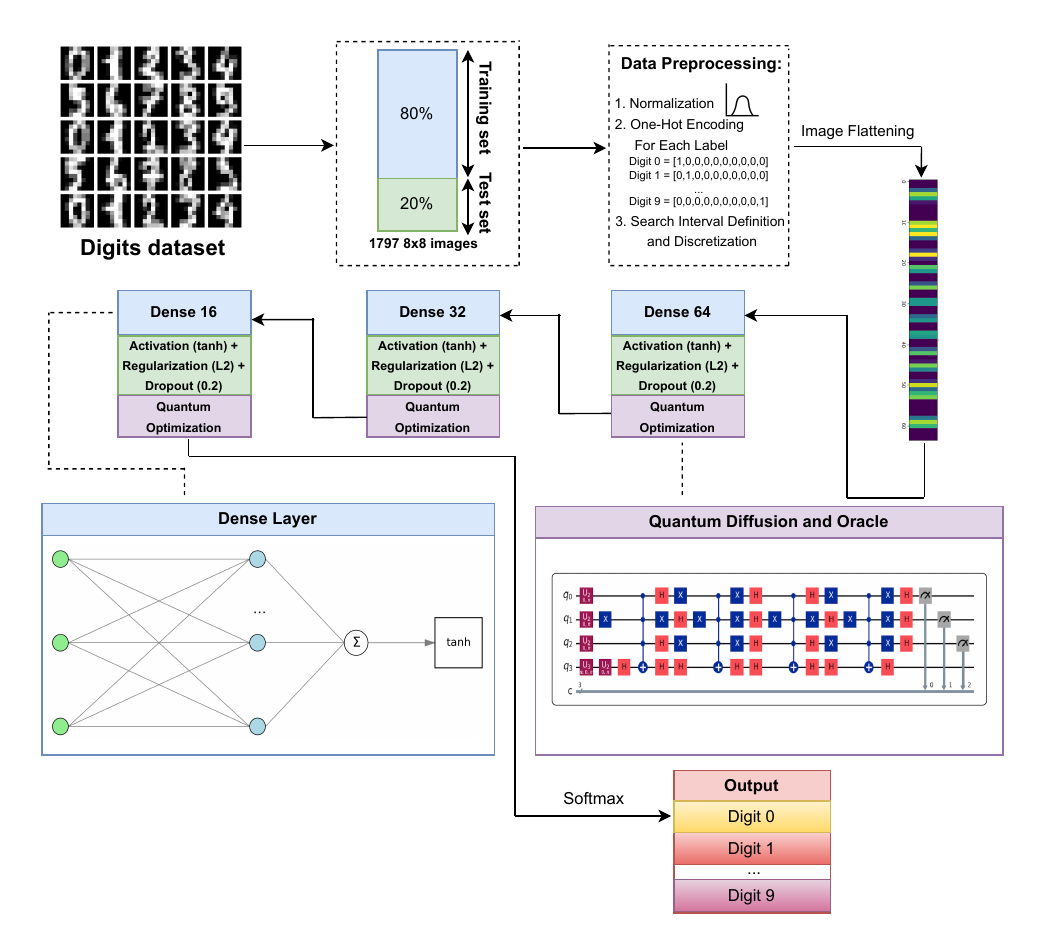}
    \caption{Our proposed method's architecture, which can be extended to an arbitrary number of hidden layers. The dataset is split into $80\%$ for training and $20\%$ for testing. Then normalization, one-hot encoding, search interval definition, and discretization are performed. The images are flattened into a one-dimensional array and fed into a hidden (dense) layer of 64 units. The hidden layer sends \emph{candidate weights} to the Grover circuit, and the circuit returns \emph{updated weights} via amplitude amplification, which is then integrated into the next hidden layer, which is made of $32$ units. The process is then repeated for the final $16$ nodes layer, and softmax is applied to determine the network's outputs.}
    \label{fig:doublecol_figure}
\end{figure*}

The main contribution of our paper is to replace the traditional backpropagation algorithm used to update the network's weights with a quantum optimization method. Figure \ref{fig:doublecol_figure} presents an overview of our proposed architecture. Consequently, in the remainder of this section, we describe the steps of the proposed algorithm for one epoch.

To control the size of the search space for each weight during optimization, we use the parameter $\alpha$. This way, we generate candidate values for each weight $w$ in a given layer by constructing an interval centered on $w$, defined as $I=[w-\alpha\sigma,w+\alpha\sigma]$, where $w$ is the current weight value, $\sigma$ is the standard deviation of the weights in the corresponding layer and $\alpha$ scales the size of the interval. The candidate values are \emph{centered around} the current weight $w$, i.e., the interval's midpoint is precisely $w$, with the candidates symmetrically distributed around it. The standard deviation $\sigma$ plays a crucial role in the adaptation of the search space. When the weights of the layer are volatile, $\sigma$ is large, leading to larger $I$. This, in turn, enables the algorithm to search a wider range of candidate values. When the weights are similar (i.e., $\sigma$ is small), the interval becomes narrow, and this focuses the search more closely to $w$. The parameter $\alpha$ is dynamically controlled according to the behavior of the loss function. If the loss decreases, the present weights and the candidate values near $w$ are good. Hence, reducing $\alpha$ (and therefore the interval) will help refine the search to fine-tune the weights. However, if the loss does not decrease, the present weight $w$ could be suboptimal and the immediate neighborhood does not offer any improvement. Consequently, when increasing $\alpha$, the search space will expand, so our method can search for a wider range of candidate values, generating a better solution. 

A function $f:\mathbb{R} \rightarrow \mathbb{R}$ is said to be Lipschitz continuous on a domain $D$ if $\exists$ a Lipschitz constant $K>0, K\in\mathbb{R}$, such that $\forall$ 2 points $x,y \in D$, we have 
\begin{equation}
\Vert f(x)-f(y)\Vert \leq K\cdot\Vert x-y\Vert,
\label{eq:lipschitz}
\end{equation}
where $\Vert\cdot\Vert$ is the 2-norm (the Euclidean norm) \cite{b15}. That means that no matter how close $x$ and $y$ are, the difference in the function values is controlled by $K$. Because our loss function is Lipschitz continuous in the proposed compact interval $I$ \cite{b42}, if the candidate is close to $w$, that is, $\vert w_{candidate} - w\vert$ is small, then the change in loss is also small and controlled; this means that we can reliably predict how the loss may vary in a small neighborhood of $w$, making the candidate search meaningful and stable. Therefore, when we narrow the interval by reducing $\alpha$, the loss improvements will be incremental and predictable, ensuring that we converge to a local minimum without the risk of a sudden change that could derail the search. If no candidate in the current neighborhood improves the loss, then $\alpha$ increases and the search space is widened. Due to Lipschitz's continuity, the loss function does not have sudden jumps, owing to the limitations imposed by $K$. With the loss function's gradient bounded, the loss difference for any small update is limited by $K$ times the change in weight; this allows our dynamic adjustment of the search space to work reliably without explicitly determining the derivative of the loss function.

Next, we present a formal description of the steps of our optimization method. 

\vspace{0.2cm}
\noindent\textbf{Quantum-Enhanced Weight Optimization (QEWO)}

\noindent \textbf{1. Initialize the network parameters.}
\begin{itemize}
  \item For the first epoch, initialize the weight matrices with values drawn uniformly from the interval \([-1,1]\):
  \[
  W^{(1)} \in \mathbb{R}^{n_1 \times n_2}, W^{(2)} \in \mathbb{R}^{n_2 \times n_3}, ..., W^{(n-1)} \in \mathbb{R}^{(n-1) \times n}
  \]
\end{itemize}

\noindent \textbf{2. Generate candidate weights for each layer.}
\begin{enumerate}
  \item For each layer \(t \in \{1,2,...,n\}\), where $n$ is the number of layers and for each weight element \(w_{ij}^{(t)}\) in \(W^{(t)}\):
  \begin{enumerate}[label=\alph*.]
    \item Calculate the standard deviation \(\sigma^{t}\) for the layer $t$.
    \item Define the search interval:
      \[
      I_{ij} := \Big[w_{ij}^{(t)} - \alpha^t\,\sigma^{t},\; w_{ij}^{(t)} + \alpha^t\,\sigma^{t}\Big].
      \]
    \item Discretize the interval \(I_{ij}\) into \(N\) candidate weights:
      \[
      x_{ij,k}^{(t)} := w_{ij}^{(t)} - \alpha^t\,\sigma^{t} + k\,\Delta x_{ij}\mbox{, for } k = 0,1,\ldots,N-1,
      \]
      where
      \[
      \Delta x_{ij} = \frac{2\,\alpha^t\,\sigma^{t}}{N-1}.
      \]
  \end{enumerate}
\end{enumerate}

\noindent \textbf{3. Evaluate the candidate weights.}
\begin{enumerate}
  \item For each candidate weight \(x_{ij,k}^{(t)}\) (the $k$-th candidate value for the weight $w_{ij}^{(t)}$):
\begin{enumerate}[label=\alph*.]
    \item Perform the forward pass:
    \[
    \begin{aligned}
    h^{(1)} &:= \phi\Big(W^{(1)}X\Big),\\[1mm]
    h^{(2)} &:= \phi\Big(W^{(2)}h^{(1)}\Big),\\[1mm]
            &\quad \vdots\\[1mm]
    h^{(n-1)} &:= \phi\Big(W^{(n-1)}h^{(n-2)}\Big),\\[1mm]
    \hat{Y} &:= \mathrm{softmax}\Big(W^{(n)}h^{(n-1)}\Big).
    \end{aligned}
    \]
    where $\phi$ is an activation function (e.g., $\tanh$, \textit{ReLU}, \textit{GeLU}, \textit{Swish}), $h^{(j)}$ represents the activation vector of the $j$-th hidden layer in the neural network, and $\hat{Y}$ denotes the network’s predicted output.
    \item Compute the loss \(L\big(x_{ij,k}^{(t)}\big)\) (e.g., using the cross-entropy loss increased with $L_2$ regularization).
  \end{enumerate}
\end{enumerate}

\noindent \textbf{4. Select promising candidates.}
\begin{itemize}
  \item Define the set of candidate indices:
    \[
    \mathcal{M} := \left\{\, k \;\vert\; L\left(x_{ij,k}^{(t)}\right) \le \min(L) + \tau \right\},
    \]
    \[\quad \tau := \texttt{tol\_ratio} \cdot \min(L)\]
    where \(\min(L)\) is the smallest loss among all candidates; initially, \(\min(L)\) is set to a given value (e.g., 5). The \texttt{tol\_ratio} parameter defines a tolerance margin around the minimum loss value found among candidate weights. Instead of selecting only the candidate with the absolute minimum loss, \texttt{tol\_ratio} allows any candidate whose loss is within \texttt{tol\_ratio} × minimum\textunderscore loss of the best candidate to be considered. This makes the search more robust to noise and grid discretization effects, improving the chance of selecting an optimal weight even if the grid does not capture the exact minimum.
\end{itemize}

\noindent \textbf{5. Optimize using Grover’s Algorithm.}
\begin{itemize}
  \item Apply Grover’s algorithm to the candidates in \(\mathcal{M}\) to identify the optimal candidate index \(k^*\) such that \[
    L\big(x_{ij,k^*}^{(t)}\big) := \min(L).
    \]
\end{itemize}

\noindent \textbf{6. Update the network weights.}
\begin{itemize}
  \item Set the weight \(w_{ij}^{(t)}\) to the selected candidate value:
    \[
    w_{ij}^{(t)} \gets x_{ij,k^*}^{(t)},
    \]
  \item Update the current minimum loss:
    \[
    \min(L) \gets L\big(x_{ij,k^*}^{(t)}\big).
    \]
\end{itemize}

\noindent \textbf{7. Adjust the scaling factor \(\alpha^t\).}\\
Let $\gamma_{down}<1$ and $\gamma_{up}>1$ be adjustment factors and $\alpha_{min}$ and $\alpha_{max}$ bounds. $\alpha_{min}$ prevents the search interval from becoming too narrow, and $\alpha_{max}$ prevents it from becoming too wide,
\[
\alpha^t \gets
\begin{cases}
\max\Big(\gamma_{down}\,\cdot\alpha^t,\,\alpha_{\min}\Big) & \text{if the loss decreases},\\[1mm]
\min\Big(\gamma_{up}\,\cdot\alpha^t,\,\alpha_{\max}\Big) & \text{otherwise}.
\end{cases}
\]

\noindent \textbf{8. Return the optimized weights.}
\[
W^{(1)},\;W^{(2)},\;W^{(3)},\;...\;,W^{(n)}.
\]

In QEWO, each weight is updated by discretizing its update range into $N$ candidate values, therefore allowing us to find the optimal candidate in $\mathcal{O}(\sqrt{N})$ steps, compared to $\mathcal{O}(N)$ steps in the classical method. Considering $M$ as the total number of weights in the network, $E$ the total number of epochs, and $T_{eval}$ the time to perform a forward pass and a loss evaluation for a single candidate, the computational cost to update all the weights from a mini-batch is $\mathcal{O}(E\cdot M \cdot \sqrt{N} \cdot T_{eval})$. Similarly, the total complexity of ADAM-based mini-batch gradient descent is $\mathcal{O}(E\cdot M \cdot T_{grad})$ \cite{b46}, where $T_{grad}$ is the time required for loss evaluation and a complete forward and backward pass (i.e., the gradient computation). Although both methods use mini-batch processing, our approach's $T_{eval}$ represents only the cost of a forward pass and loss computation, as it does not include backpropagation; this reduces the per-candidate evaluation time. We know that $N$ denotes the number of quantum states, and $N=2^n$, where $n$ is the number of bits representing each value, so the candidate search space grows exponentially with $n$. However, Grover's algorithm reduces the number of iterations to $2^{\frac{n}{2}}$ . In QEWO, because we work with small and moderate candidate grid sizes, we avoid the otherwise explosive growth in the search space while still benefiting from the quadratic reduction, meaning that our discrete search is efficient and effective for moderate $N$. If $N$ were allowed to grow too large, even with the quadratic speedup, the absolute number of evaluations would still become high, and the candidate values might be so spread out that the precision needed to capture the optimal weight is lost. This careful choice of candidate grid size (parametrized with \texttt{tol\_ratio}) leverages Grover's quadratic speedup while maintaining the precision necessary for effective weight optimization. All these make our approach especially well-suited for practical applications where both efficiency and accuracy are critical.

\section{Experimental results}
\label{sec:exp}
\subsection{Exerimental Setup}
We applied the quantum optimization method on an MLP using the \textit{Wine} and \textit{Digits} datasets for training and testing. The \textit{Wine} dataset contains $178$ samples of wine, divided into three classes ($59$ in class $1$, $71$ in class $2$, and $48$ in class $3$). The samples are characterized by $13$ numerical features (alcohol content, color, proline concentration, and more) \cite{b27}. We split the dataset into a training set (80\% of the actual dataset) and a testing set (20\% of the actual dataset) to ensure that the model can generalize from previously unseen, novel data. The \textit{Digits} dataset dataset contains 1797 8x8 hand-written images, with digits from $0$ to $9$ \cite{b45}. Like the \textit{Wine} dataset, the \textit{Digits} dataset was also split into a training set and a testing set, containing 80\% of the data and 20\% of the data, respectively. We also trained a conventional MLP using the backpropagation algorithm with mini-batch gradient descent and ADAM optimizer for comparison.
All networks were trained for $10$ epochs under different conditions (mentioned below), having a dropout with probability $p=0.2$ and $L_2$ regularization using the Frobenius norm; the activation function is the hyperbolic tangent, except for the $4^{th}$ experiment, and softmax is used for classification. For the hidden layers, we set \texttt{tol\_ratio} = $0.05$, and for the output layer, \texttt{tol\_ratio} = $0.1$; we chose these values based on empirical testing, to make a trade-off between strictness and flexibility. A \texttt{tol\_ratio} of $0.05$ for the hidden layers provides a relatively narrow margin; this ensures that only candidates very close to the best-known loss value are considered, which helps maintain a sharp distinction between near-optimal and suboptimal weight configurations. We use a \texttt{tol\_ratio} of $0.1$ for the output layer to allow a slightly larger margin. This adjustment accounts for the higher sensitivity of the final output distribution to weight variations. The larger tolerance at this stage helps ensure that small fluctuations do not overly limit the candidate set, thereby aiding the discrete search process in effectively exploring the weight space. 

Our current experiments rely on classical simulations of our quantum weight optimization, where the candidate grid size grows exponentially with the number of qubits (since $N=2^n$). This exponential growth dramatically increases memory and computational burden, making it infeasible to simulate on larger datasets like MNIST with available resources. However, this limitation is intrinsic to simulating quantum systems on classical computers. In the future, when fully functional quantum hardware becomes available---with the capacity to handle large numbers of qubits natively---our approach will no longer be hampered by these classical resource constraints. 


\subsection{Experiment 1: Small Neural Network on a Small Dataset}

\begin{figure}[H]
\begin{center}
  \includegraphics[scale=0.25, width=0.8\columnwidth]{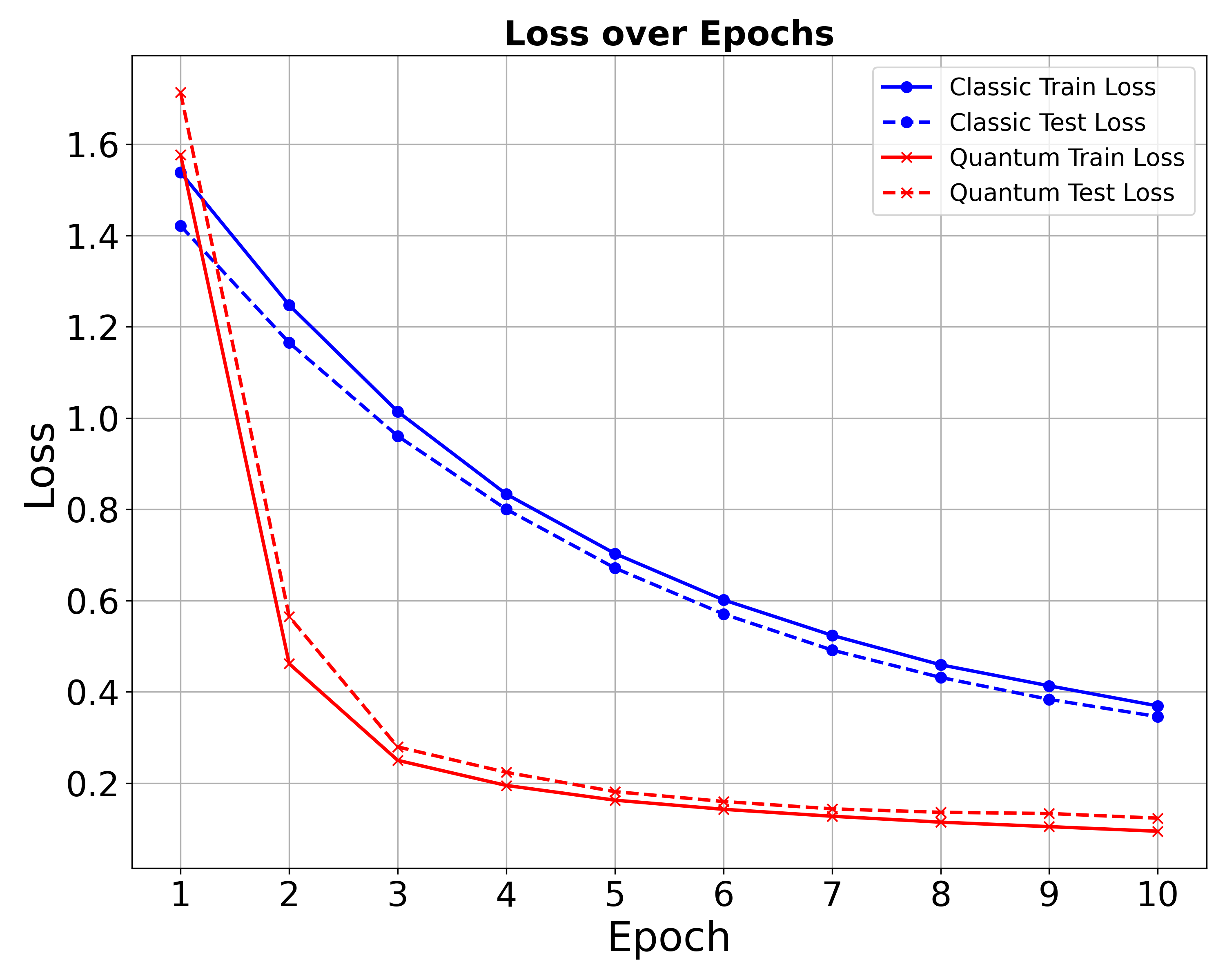}\\[2pt] 
  \caption{
    Evolution of training and test loss over 10 epochs, comparing classical ADAM (blue) and quantum Grover-optimized (red) neural networks. 
    Circle markers refer to training loss, while dashed lines indicate test loss. 
    The quantum-optimized approach acquires faster convergence and achieves a lower final loss than the classical method.}
  
\label{fig:1}
\end{center}
\end{figure}

Here, both the quantum-optimized and classical networks share the same architecture: an input layer with $13$ units (one per feature), one hidden layer with $32$ units, and an output layer with $3$ neurons (one per class). The quantum-optimized MLP uses $32$ candidate values (the number of discrete values found within the search space) for the hidden layer and $64$ for the output layer, with an initial search factor $\alpha = 0.1$, $\gamma_{down} = 0.95$, and $\gamma_{up} = 1.05$. Weights are randomly initialized from a uniform distribution in $[-1, 1]$. The model was trained $10$ times, and the best-performing run was retained for the plots and tables, although all runs exhibited very similar results in terms of convergence and accuracy. Our experiments showed that the loss function decreased abruptly for our Grover optimization as the number of epochs increased, thus proving that our quantum-optimized network can converge to optimal parameters very quickly, as shown in Figure \ref{fig:1}. We also notice that the classical ADAM-based network has a lower loss at the very first epoch, as shown in Tables \ref{tab:train_loss}, \ref{tab:test_loss1}, and Figure \ref{fig:1}; this is because the quantum approach randomly selects the initial search space for the weights and, therefore, the starting weights may be placed very far from the optimum. However, after one epoch, the quantum-optimized network rapidly approaches an optimal solution, showing its efficiency in identifying the best weights, as presented in Table \ref{tab:train_loss}, Table \ref{tab:test_loss1} and Figure \ref{fig:1}. Furthermore, the ADAM optimizer usually needs more iterations to converge to the minimum loss than the quantum optimizer; this suggests that Grover’s algorithm, which underpins our quantum-optimized method, needs fewer epochs and less training time to find the best parameters, a clear advantage over the classical approach. 

\begin{figure}[H]
\begin{center}
\includegraphics[scale=0.25,width=0.8\columnwidth]{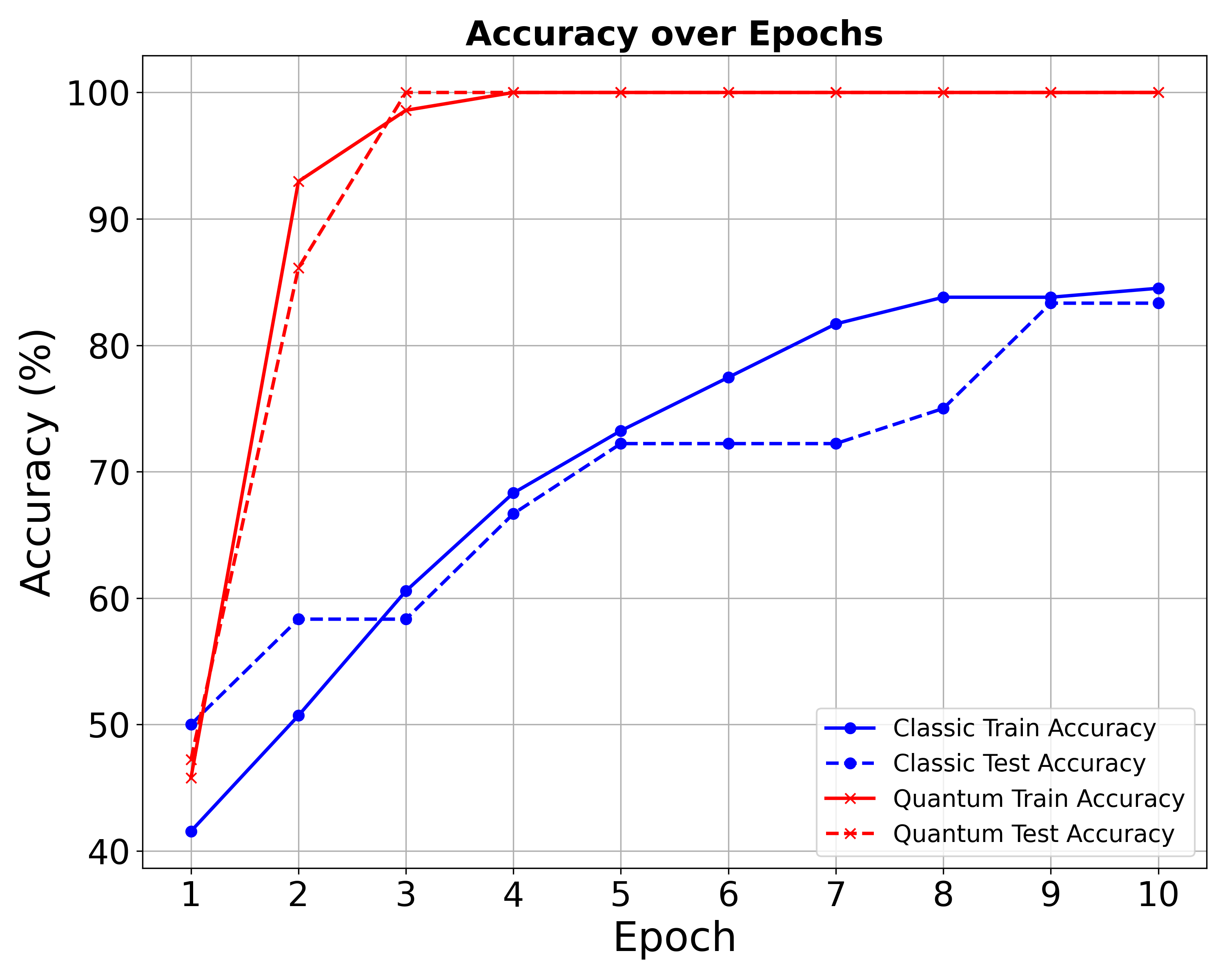}
 \caption{
   Comparison between training and improvement in test accuracy over 10 epochs. The accuracy is represented by dashed lines for test and circle markers for training. The quantum-optimized model is able to learn the data structure very quickly and achieve near-optimal accuracy within the first few epochs of training, while the classical approach has a more gradual improvement in accuracy.
}
\label{fig:2}
\end{center}
\end{figure}

Figure \ref{fig:2} shows the evolution we obtain for accuracy over the epochs, with the data derived from Tables \ref{tab:test_loss2} and \ref{tab:test_accuracy}. From the accuracy plot in Figure \ref{fig:2}, we see that the training and test accuracy of the quantum-optimized network reaches $100$\% within the first few epochs; this shows that the network converges quickly to the optimal solution, indicating the effectiveness of the training data in capturing data patterns. The faster optimization of the quantum-optimized model also implies that Grover’s algorithm is able to identify good weights more effectively, thus reducing training time while achieving perfect classification in both training and test sets.

\begin{table}[H]
    \scriptsize
    \centering
    \caption{Train loss comparision between classic (ADAM) and quantum (Grover-based) optimization}
    \begin{tabular}{c c c c c}
        \textbf{Epoch} & \textbf{Classic} & \textbf{Quantum} & \textbf{Reduction} & \textbf{Improvement (\%)} \\ 
        & \textbf{Train Loss} & \textbf{Train Loss} &  &  \\ 
        1  & 1.5384  & 1.5770  & -0.0386  & -2.51\%  \\
        2  & 1.2477  & 0.4616  & 0.7861   & 63.00\%  \\
        3  & 1.0140  & 0.2498  & 0.7642   & 75.38\%  \\
        4  & 0.8332  & 0.1948  & 0.6384   & 76.63\%  \\
        5  & 0.7030  & 0.1625  & 0.5405   & 76.90\%  \\
        6  & 0.6015  & 0.1425  & 0.4590   & 76.30\%  \\
        7  & 0.5238  & 0.1273  & 0.3965   & 75.70\%  \\
        8  & 0.4592  & 0.1143  & 0.3449   & 75.12\%  \\
        9  & 0.4131  & 0.1045  & 0.3086   & 74.70\%  \\
        10 & 0.3691  & 0.0944  & 0.2747   & 74.41\%  \\ 
    \end{tabular}
    \label{tab:train_loss}
\end{table}

\begin{table}[H]
    \scriptsize
    \centering
    \caption{Test loss comparision between classic (ADAM) and quantum (Grover-based) optimization}
    \begin{tabular}{c c c c c}
        \textbf{Epoch} & \textbf{Classic} & \textbf{Quantum} & \textbf{Reduction} & \textbf{Improvement (\%)} \\ 
        & \textbf{Test Loss} & \textbf{Test Loss} &  &  \\ 
        1  & 1.4213  & 1.7135  & -0.2922  & -20.56\%  \\
        2  & 1.1652  & 0.5650  & 0.6002   & 51.50\%  \\
        3  & 0.9606  & 0.2792  & 0.6814   & 70.90\%  \\
        4  & 0.8001  & 0.2235  & 0.5766   & 72.07\%  \\
        5  & 0.6713  & 0.1812  & 0.4901   & 72.73\%  \\
        6  & 0.5705  & 0.1596  & 0.4109   & 71.82\%  \\
        7  & 0.4915  & 0.1435  & 0.3480   & 70.86\%  \\
        8  & 0.4315  & 0.1359  & 0.2956   & 68.57\%  \\
        9  & 0.3834  & 0.1333  & 0.2501   & 65.22\%  \\
        10 & 0.3456  & 0.1230  & 0.2226   & 64.38\%  \\ 
    \end{tabular}
    \label{tab:test_loss1}
\end{table}

\begin{table}[H]
    \centering
    \scriptsize
    \caption{Train accuracy comparision between classic (ADAM) and quantum (Grover-based) optimization}
    \begin{tabular}{c c c c c}
        \textbf{Epoch} & \textbf{Classic} & \textbf{Quantum} & \textbf{Improvement} & \textbf{Improvement (\%)} \\ 
        & \textbf{Train Acc \%} & \textbf{Train Acc \%} &  &  \\ 
        1  & 41.55  & 45.77  & 4.22  & 10.15\%  \\
        2  & 50.70  & 92.96  & 42.26   & 83.34\%  \\
        3  & 60.56  & 98.59  & 38.03   & 62.79\%  \\
        4  & 68.31  & 100.00  & 31.69   & 46.41\%  \\
        5  & 73.24  & 100.00  & 26.76   & 36.53\%  \\
        6  & 77.46  & 100.00  & 22.54   & 29.10\%  \\
        7  & 81.69  & 100.00  & 18.31   & 22.41\%  \\
        8  & 83.80  & 100.00  & 16.20   & 19.34\%  \\
        9  & 83.80  & 100.00  & 16.20   & 19.34\%  \\
        10 & 84.51  & 100.00  & 15.49  & 18.33\%  \\ 
    \end{tabular}
    
    \label{tab:test_loss2}
\end{table}

\begin{table}[H]
    \scriptsize
    \centering
    \caption{Test accuracy comparision between classic (ADAM) and quantum (Grover-based) optimization}
    \begin{tabular}{c c c c c}
        \textbf{Epoch} & \textbf{Classic} & \textbf{Quantum} & \textbf{Improvement} & \textbf{Improvement (\%)} \\ 
        & \textbf{Test Acc \%} & \textbf{Test Acc \%} &  &  \\ 
        1  & 50.00  & 47.22  & -2.78  & -5.56\%  \\
        2  & 58.33  & 86.11  & 27.78  & 47.62\%  \\
        3  & 58.33  & 100.00  & 41.67  & 71.43\%  \\
        4  & 66.67  & 100.00  & 33.33  & 50.00\%  \\
        5  & 72.22  & 100.00  & 27.78  & 38.58\%  \\
        6  & 72.22  & 100.00  & 27.78  & 38.58\%  \\
        7  & 72.22  & 100.00  & 27.78  & 38.58\%  \\
        8  & 75.00  & 100.00  & 25.00  & 33.33\%  \\
        9  & 83.33  & 100.00  & 16.67  & 20.00\%  \\
        10 & 83.33  & 100.00  & 16.67  & 20.00\%  \\ 
    \end{tabular}
    \label{tab:test_accuracy}
\end{table}

In addition, our algorithm performed better on the small dataset than the variational quantum method (VQC). The algorithm proposed in \cite{b36} has been trained on the \emph{Iris} dataset (which consists of 3 classes of 50 instances, thus 150 samples, each having 4 numerical features \cite{b43}, meaning that \emph{Iris} is smaller than the dataset we used). Under these conditions, the quantum variational perceptron has achieved $90 \%$ accuracy in the test data, while their Grover-accelerated version has achieved $99\%$ accuracy in the test dataset. Our method has achieved a test accuracy of $100\%$, demonstrating that our classical NN with the quantum optimized weights significantly outperforms the classically optimized variational quantum method in small datasets.

Our approach cannot be directly compared to the OQP of \cite{b28} because the OQP updates one parameter at a time, continuing until every weight is correct. As a result, the overall accuracy cannot be measured until the entire model is fully updated. In contrast, our model updates all the weights simultaneously within each epoch, which allows us to evaluate and report overall accuracy at the end of every epoch.

The soft quantum perceptron from \cite{b39} was trained on various small datasets, including the \emph{moons} dataset which consists of 100 samples with 2 features \cite{b44}, and achieved a test accuracy of $100\%$, just like our model. While the soft quantum perceptron has been tested on small, low-dimensional datasets (like the \emph{moons} dataset), our model has been evaluated on the \emph{Wine} dataset, which is larger and more complex. Achieving 100\% test accuracy in the \emph{Wine} dataset shows that our model scales effectively and handles higher-dimensional data without compromising performance. Also, their classical update via the backpropagation algorithm suffers from several disadvantages. In particular, it is highly sensitive to the choice of hyperparameters such as the learning rate, which can lead to issues such as vanishing or exploding gradients, especially in deeper networks or in the presence of complex, high-dimensional data. This sensitivity often requires extensive tuning and may result in suboptimal convergence. In contrast, our model employs a quantum-enhanced optimization strategy that leverages quantum amplitude amplification, allowing for a more robust global search for optimal weights. Our approach accelerates convergence and scales more effectively to complex datasets, as evidenced by our 100\% test accuracy on the larger, higher-dimensional \emph{Wine} dataset.

\subsection{Experiment 2: Deep Neural Network on a Larger Dataset}

\begin{figure}[H]
\begin{center}
\includegraphics[scale=0.25,width=0.8\columnwidth]{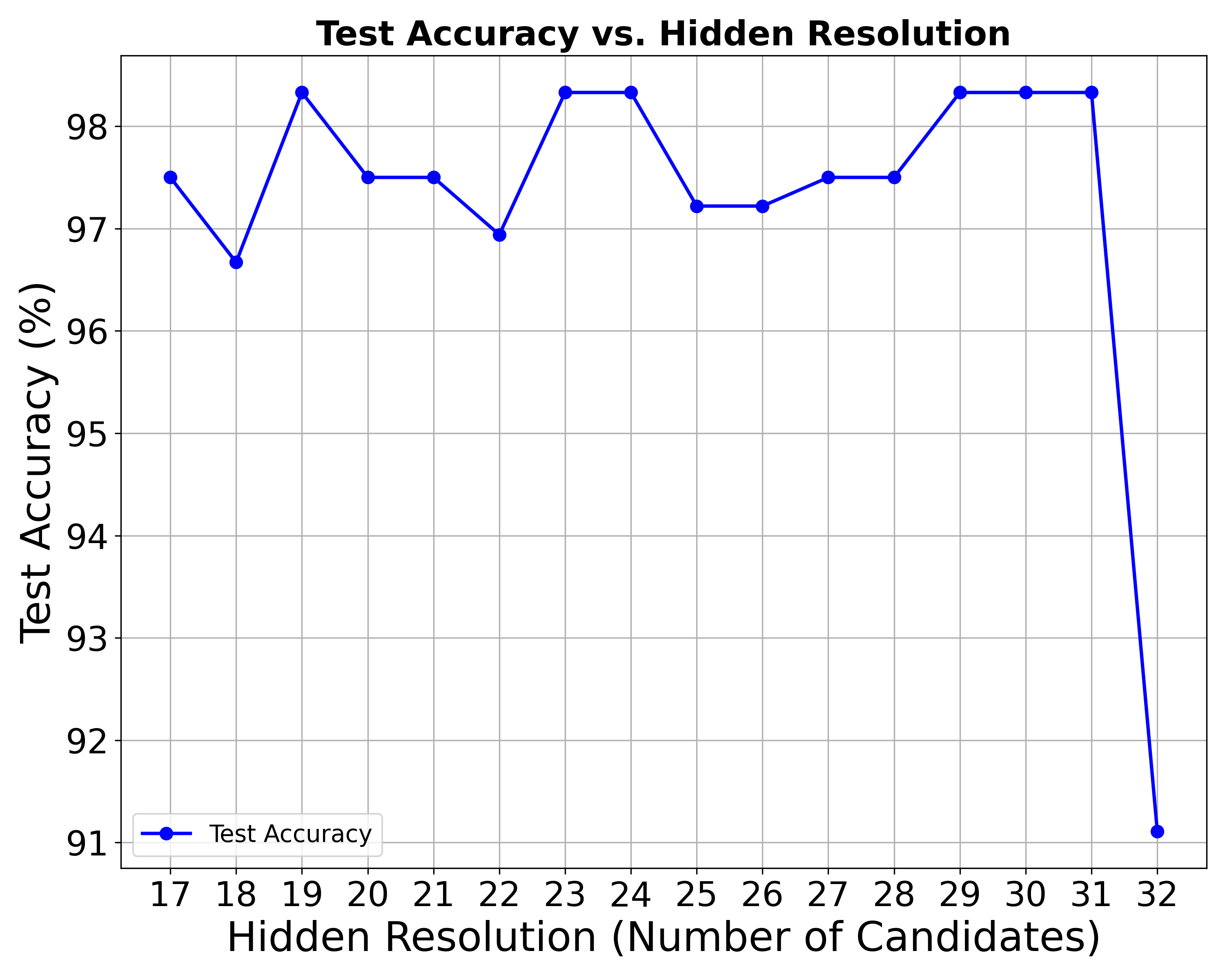}
\caption{
   Illustration on how the final test accuracy changes as we vary the hidden resolution (i.e., the number of candidate values for each layer) from 17 to 32. For resolutions between 17 and 31, the test accuracy remains consistently high—around 97\% or more—indicating stable and robust performance. However, at resolution 32, the accuracy suddenly drops to approximately 91\%, suggesting that excessively fine discretization of the candidate space can lead to numerical instability or degraded model performance.
}
\label{fig:3}
\end{center}
\end{figure}

\begin{figure}[H]
\begin{center}
\includegraphics[scale=0.25,width=0.8\columnwidth]{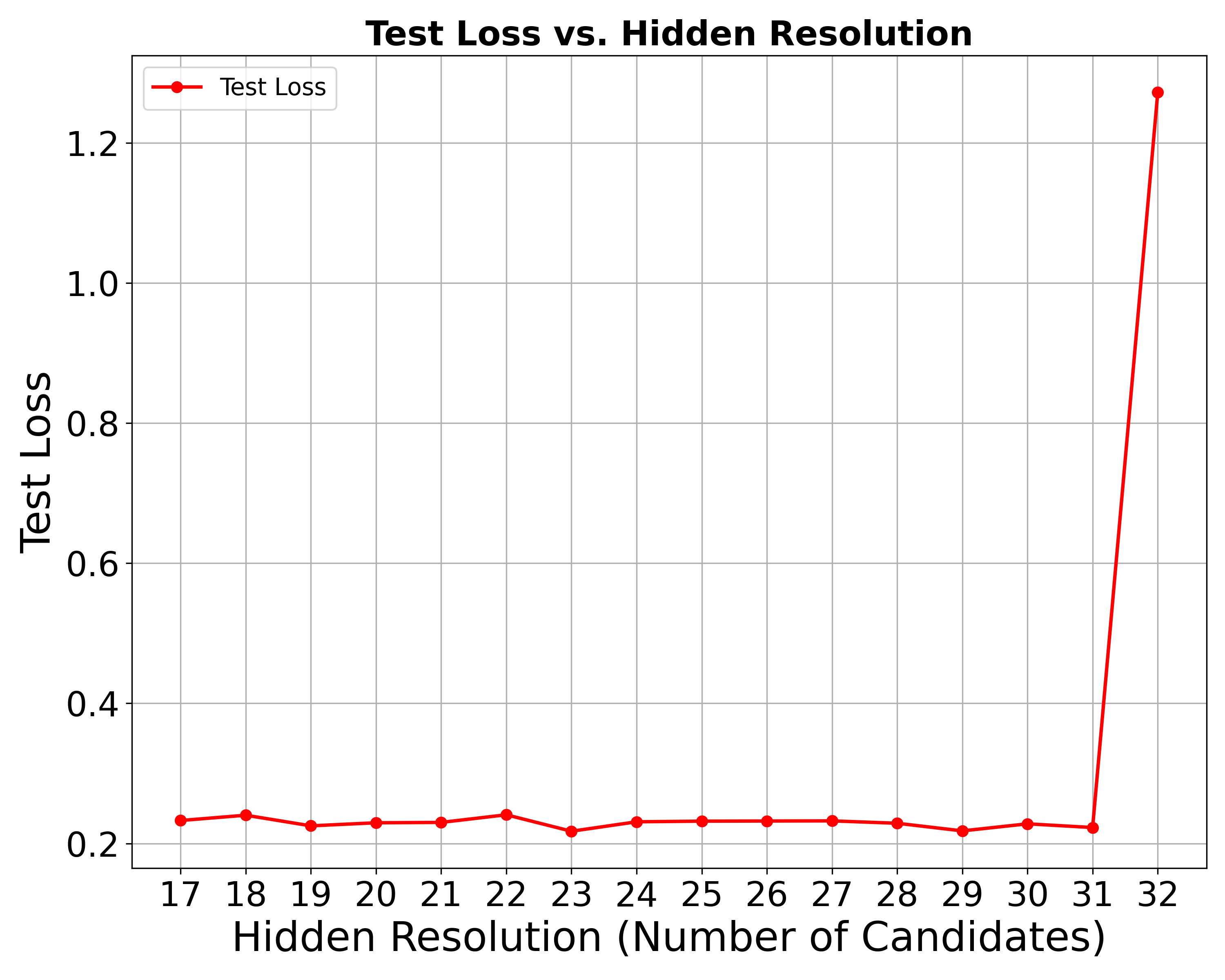}
\caption{
   Illustration on how the final test loss evolves as the hidden resolution (i.e., the number of candidate values for each layer) increases from 17 to 32. For resolutions between 17 and 31, the test loss remains consistently low, indicating stable training and strong generalization. However, at resolution 32, the test loss spikes sharply to around 1.27, suggesting that an overly fine discretization of the weight space can destabilize the optimization process and degrade the model’s performance, namely overfitting. 
}
\label{fig:4}
\end{center}
\end{figure}

One of the main contributions of our work is its scalability. Indeed, most QNNs are trained on small datasets. However, we could make our network deep: 3 hidden layers with 64, 32, and 16 units, respectively. We have trained our network using the scikit-learn \emph{Digits} dataset, using random weight initialization, with values in the interval $[-1 , 1]$. We trained the network using the same adjustment factors as in the previous experiment; however, we modified the number of potential candidates for each hidden layer, starting from 17 to 32, while keeping the output layer constant with 64 values. Having 3 hidden layers, each layer's resolution space takes the same values, from 17 to 32, in multiple iterations. We noticed that the hidden resolution (i.e., from 17 to 31) maintains a high accuracy and a decreasing loss function. However, when we set the number of search candidates at 32, the precision decreased, thus showing that a large number of candidates (that is, more than a certain threshold) affects the performance of the model, as shown in Figures \ref{fig:3} and \ref{fig:4}.

\begin{figure}[H]
\begin{center}
\includegraphics[scale=0.25,width=0.8\columnwidth]{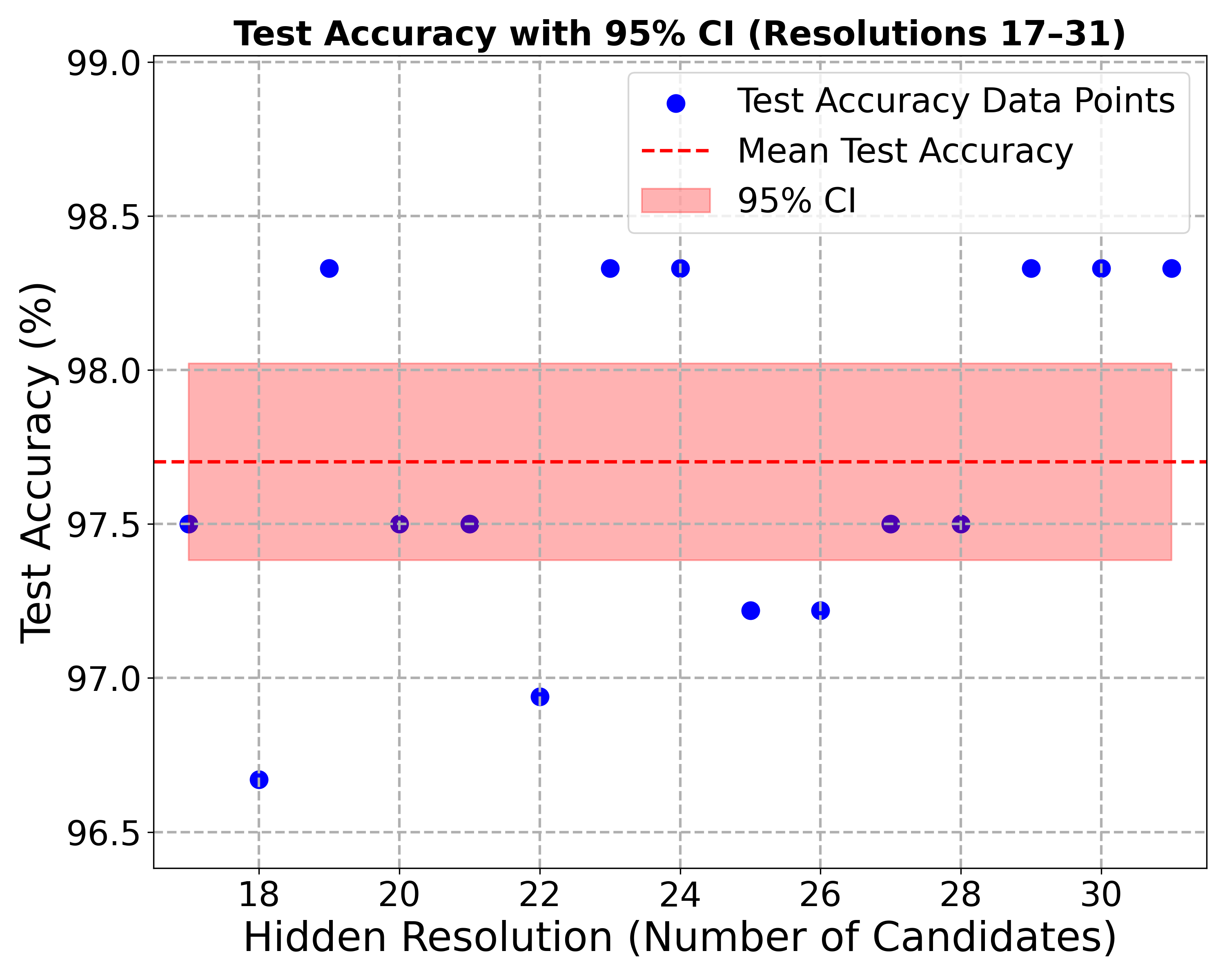}
\caption{
   The final test accuracies (blue dots) achieved by varying the hidden resolution from 17 to 31, along with the mean accuracy (dashed red line) and the 95\% confidence interval (shaded region) around that mean. The tight clustering of data points within the shaded area indicates both high and consistent performance across this range of hidden resolutions.
}
\label{fig:5}
\end{center}
\end{figure}

The mean accuracy for the 15 experiments (corresponding to resolution values from 17 to 31) observed on the deep network is $\Tilde{x}\approx 97.70 \%$. The estimated standard deviation is $ 0.56\%$ after computing the deviations from the mean for each data point. For $n=15$ data points, using the statistical Student's $t$-test, the appropriate $t$ value for a $95\%$ confidence interval is roughly
$t_{0.025,14}\approx 2.14$. The margin of error (ME) is given by

\begin{equation}
ME=t\times \frac{\sigma}{\sqrt{n}}=2.14\cdot \frac{0.56}{\sqrt{15}}\approx 0.31 \% .
\label{eq:2}
\end{equation}
Henceforth, the $95\%$ confidence interval for the mean test accuracy, according to Figure \ref{fig:5}, is
\begin{equation}
[\Tilde{x}-ME,\Tilde{x}+ME]=[97.39\%,98.01\%].
\label{eq:3}
\end{equation}

\begin{figure}[H]
\begin{center}
\includegraphics[scale=0.25,width=0.8\columnwidth]{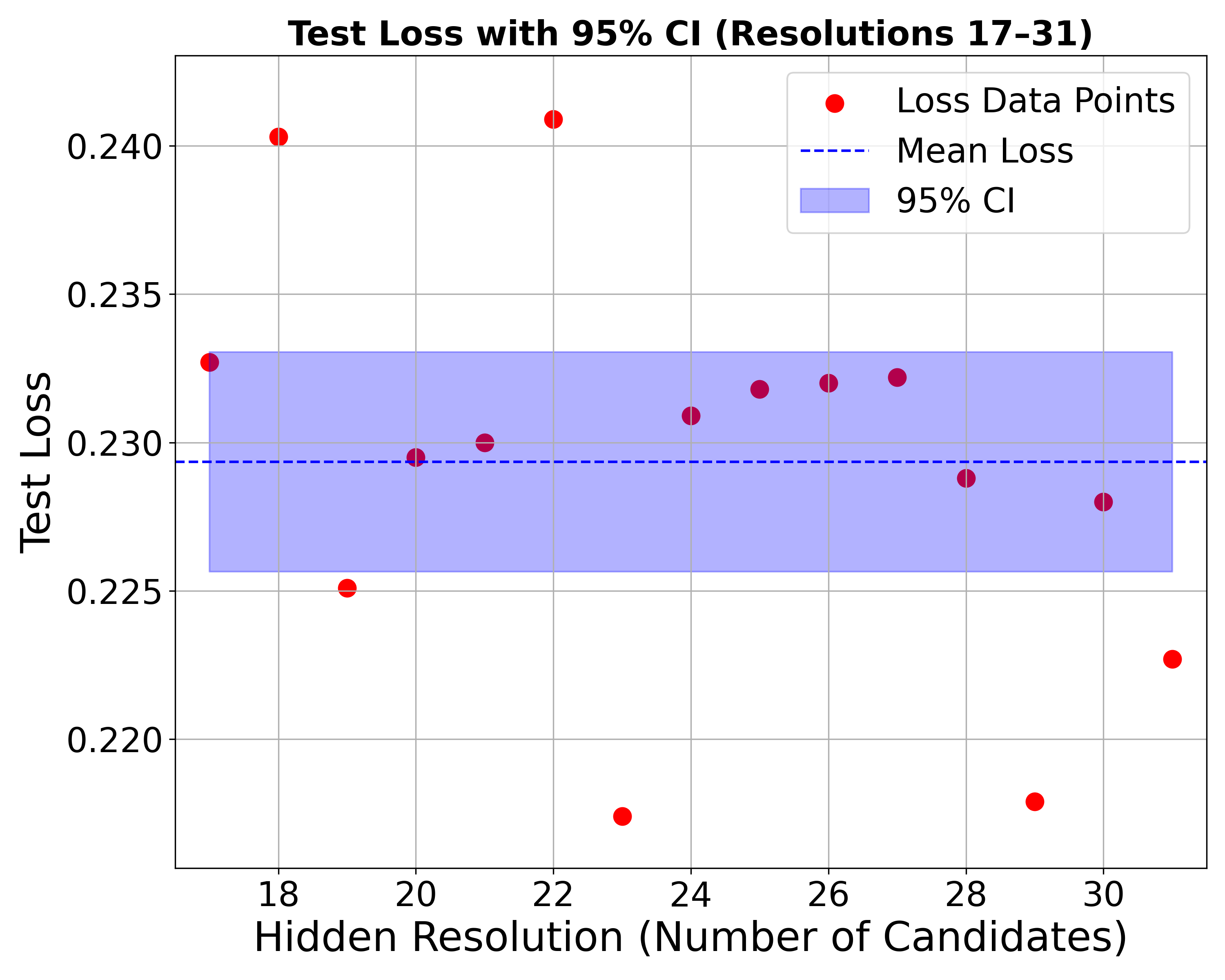}
\caption{
   Illustration of final test loss values (red dots) obtained for hidden resolutions ranging from 17 to 31. The dashed blue line represents the average loss across all resolutions, and the shaded region indicates the 95\% confidence interval around that mean. The narrow interval and consistently low loss values highlight the model’s stable performance in this resolution range.
}
\label{fig:6}
\end{center}
\end{figure}

Applying the same test for the loss function, we find that the confidence interval is $[0.2256,0.2330]$, as shown in Figure \ref{fig:6}, therefore confirming that the test loss remains very low and stable for resolutions 17--31. Statistical analysis supports the conclusion that our algorithm operates best with a hidden resolution in the range of 17--31. Attempting to push resolution further to 32 harms performance, likely due to increased sensitivity or noise in the weight optimization process. By examining too many candidate weight values at once, the algorithm becomes affected by minor local fluctuations, making it harder to lock onto a truly optimal region. Consequently, large intervals increase numerical instability, increasing the sensitivity of the Grover algorithm and leading to erratic weight updates that reduce the model's convergence stability and accuracy. Moreover, very fine weight increments (due to many more possible candidate weights) can lead the learning process to small, dataset-specific patterns that do not generalize well, thus decreasing the test accuracy, even though the training loss might appear to improve or remain stable. 

As the optimization proceeds strictly layer‑by‑layer, the register must hold only the candidate weights belonging to the layer currently being tuned. For a hidden layer, 19 (the value for which the model achieves the best performance) candidate weights require $\lceil{\log_2 19}\rceil=5$ index qubits, and Grover’s routine adds a single ancilla, so the search occupies just 6 qubits. Once that layer has converged, the 6‑qubit register is reset and recycled for the next hidden layer, preventing any cumulative growth. The output layer, although larger, still fits comfortably: 64 candidate weights demand $\lceil{\log_2 64}\rceil=6$ index qubits plus one ancilla — 7 qubits in total. Because the hidden‑layer register is cleared before the output search begins, the hardware is never required to host more than these 7 logical qubits at any stage of training. This peak demand is more than 5 times lower than the roughly 40 qubits reported in \cite{b39} for a problem of comparable size.

\begin{figure}[H]
\begin{center}
\includegraphics[scale=0.25,width=0.8\columnwidth]{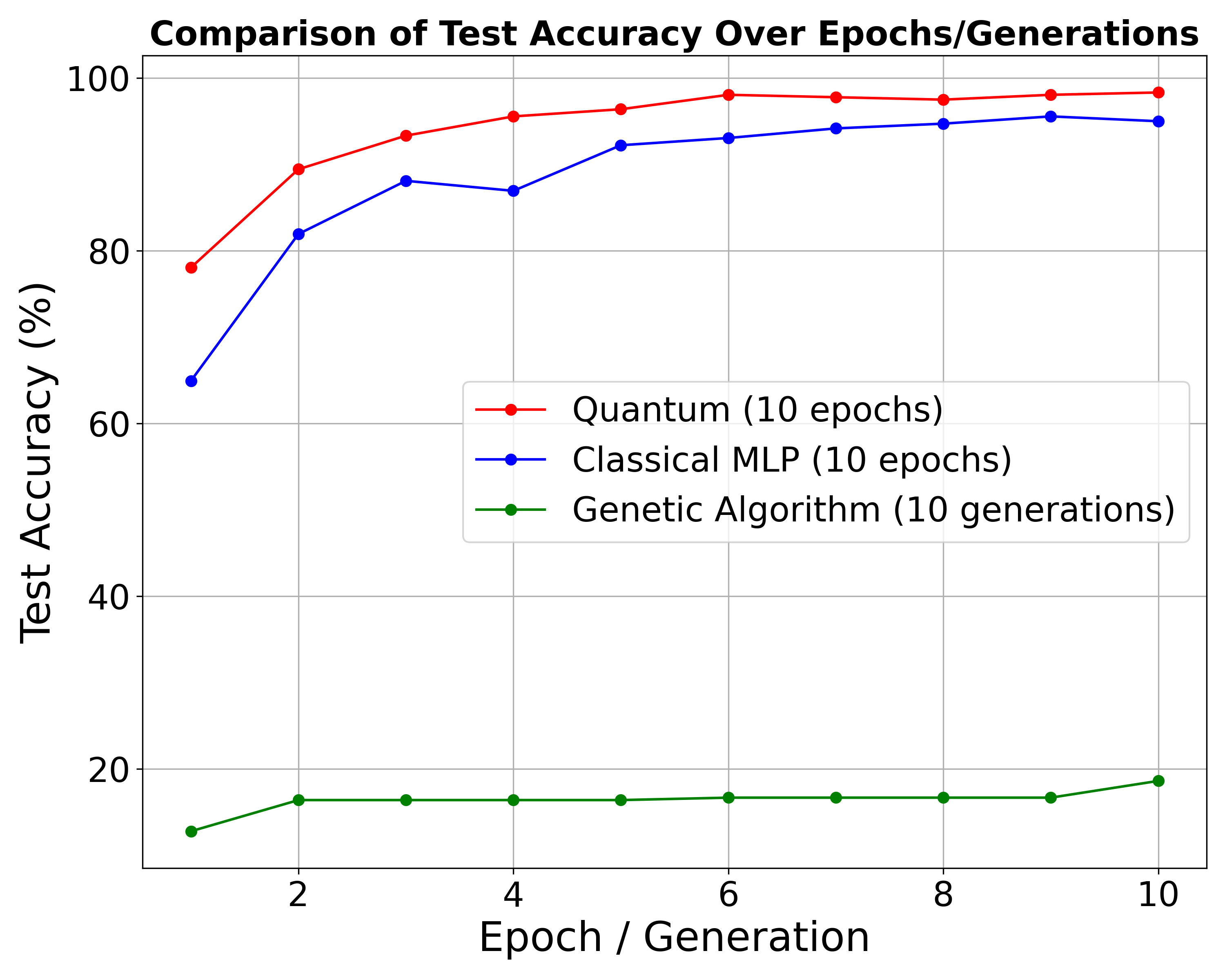}
\caption{
    Comparing the mean test accuracy per epoch across three models, each trained 10 times for 10 epochs, across their respective training procedures: a quantum model (blue, the one with $19$ candidates for each hidden layer, i.e., the best result obtained during the training process), a classical MLP (orange), and a gradient-free genetic algorithm (green). The x-axis represents the training epochs or generations, and the y-axis is the percentage of correctly classified test samples. While both the quantum and classical models rapidly converge to high accuracy, the genetic algorithm exhibits a slower climb, underscoring different optimization behaviors and convergence rates among the three approaches.
}
\label{fig:7}
\end{center}
\end{figure}

Figure \ref{fig:7} shows the evolution of accuracy over epochs for 3 approaches: the quantum weight optimization (the one with the best results for the deep NN, i.e., 19 candidate values for each hidden layer and 64 for the output), a classical one (the one with the best results after training $10$ times) and one optimized with a gradient-free genetic algorithm (using PyGad, a Python framework specially designed for handling genetic algorithms). Figure \ref{fig:7} demonstrates that while both the quantum and classical approaches converge rapidly and achieve high accuracy, the genetic algorithm shows a slower climb and settles at a noticeably lower performance level. This discrepancy underscores the differences in optimization strategies, amplitude amplification in the quantum model versus gradient training in the classical model versus population-based evolution in the genetic algorithm, and the impact these methods have on convergence behavior and final accuracy. Our quantum-optimized model demonstrates better performance when dealing with a deep neural network and a large dataset such as \emph{Digits}, and shows even stronger performance when dealing with small datasets such as \emph{Iris} or \emph{Wine}, therefore showing its scalability and capability to rival classical optimization methods and its potential of being integrated in large-scale applications.

\subsection{Experiment 3: Employing Varied Weight Initializations}
Even though our method does not use gradients, the way we initialize the weights matters. For the random initialization using a uniform distribution over $[-1, 1]$, for the same network, the weights start with a broader spread with $19$ candidate values for each hidden layer. This wider variance means that when our quantum-optimized discrete search explores candidate values, it covers a larger and more diverse region of the weight space. As a result, the search is more likely to escape local minima and find a better solution.

\begin{figure}[H]
\begin{center}
\includegraphics[scale=0.25,width=0.8\columnwidth]{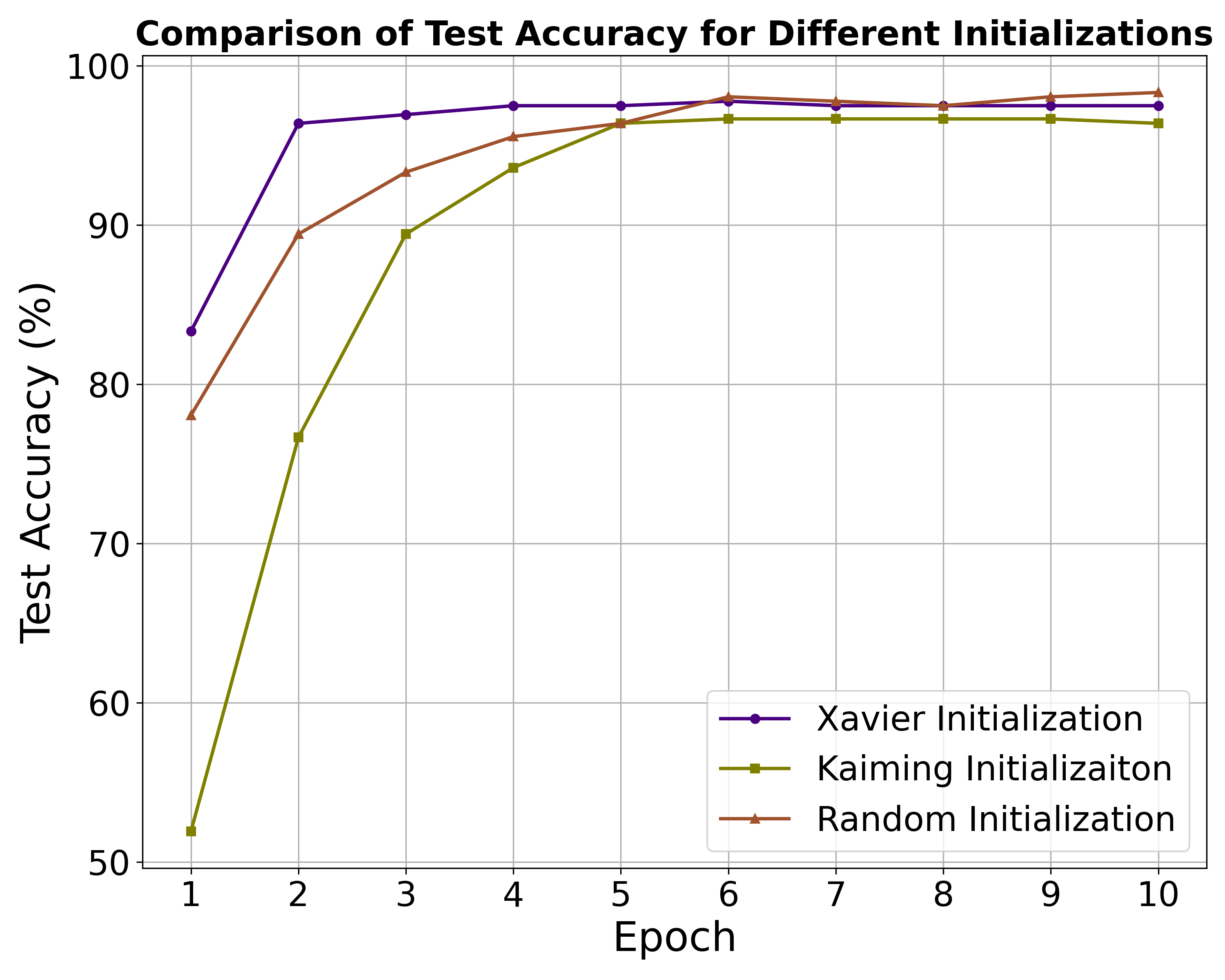}
\caption{
   Evolution of the mean test accuracy per epoch for 3 initialization strategies, averaged over 10 runs of 10 epochs each. The indigo curve (Xavier) shows moderate initial accuracy with rapid convergence above $95\%$. The olive curve (Kaiming) starts with a lower convergence but quickly improves to match high-accuracy performance by epochs 4–5. The sienna curve (Random) exhibits a high initial accuracy, reaching nearly 99\% by later epochs.
}
\label{fig:8}
\end{center}
\end{figure}

In contrast, methods such as Xavier or Kaiming typically produce weights with a smaller variance. Although this is beneficial in gradient-based training—since it helps control the variance across layers—in our gradient-free approach, it can be a drawback. A narrower initial range causes the candidate values to cluster closely together, which may limit the effectiveness of our local search. Ultimately, random initialization seems to offer a more diverse starting point, which improves the overall performance of our quantum optimization process, as shown in Figure \ref{fig:8}.

\subsection{Experiment 4: Different Activation Functions}
We trained the previously best-performing model using the GeLU and Swish activation functions. Each configuration underwent 10 independent training runs, each consisting of 10 epochs. The mean accuracy obtained over these 10 runs using $\tanh$ is $98.33\%$; when using GeLU, the mean accuracy is $97.22\%$; after Swish, it is $95.56\%$. Therefore, the hyperbolic tangent is a better option for our method.

Our approach relies on a discrete, quantum-optimized search over candidate weight values. With its steep non-linearity and clear saturation regions, the hyperbolic tangent tends to produce a loss landscape with more pronounced features or distinct regions. This rugged landscape allows our discrete search to more effectively distinguish between candidate weights and identify promising regions for optimization. In contrast, GeLU and Swish are smoother activation functions that generate a flatter loss landscape. Although these activations often benefit gradient-based training in deep networks, the smoother transitions in our gradient-free, quantum discrete search setting make it harder for the search algorithm to differentiate between candidates. As a result, the discrete quantum search struggles to find the optimal weight values when using GeLU or Swish, leading to lower overall performance.

Overall, the sharper transitions provided by the hyperbolic tangent are better aligned with our quantum discrete search approach, resulting in a more effective exploration of the weight space and superior performance in our experiments.
\subsection{Experiment 5: Noise Sensitivity}
This experiment simulated realistic NISQ (Noisy Intermediate-Scale Quantum) conditions by applying a depolarizing noise model to the quantum circuits involved in our optimization method. Specifically, every single-qubit gate was subjected to an error rate of $0.5\%$ $(p_1 = 0.005)$ and every two-qubit gate to an error rate of $2\%$ $(p_2 = 0.02)$, with both error types applied simultaneously. This setup mirrors the typical gate fidelities for current quantum hardware \cite{b47}.

Under noisy conditions, our discrete quantum search reached a  mean test accuracy of $97.22\%$, averaged over 15 runs of 10 epochs each, compared to $97.7\%$ in an ideal noise-free setting. Notably, our method still outperformed the classical MLP, which reached around $95\%$ accuracy. Although the accuracy drop is modest, it demonstrates that even moderate noise can slightly blur the loss landscape features on which our discrete search relies to differentiate between candidate weights.

Nevertheless, these results hint that, while our method exhibits promising robustness against noise, further noise mitigation or error correction will be essential when scaling to larger systems or more complex architectures.

\section{Conclusions}
\label{sec:concl}

In this work, we introduced a novel quantum-enhanced weight optimization method that replaces traditional NN gradient-based updates with a quantum-optimized discrete search procedure. By discretizing the update range of each weight into a candidate grid of $N$ values and employing Grover’s algorithm, our approach reduces the candidate evaluation from $\mathcal{O}(N)$ to $\mathcal{O}(\sqrt{N})$ per weight. When combined with mini-batch processing---where $T_{\text{eval}}$ denotes the cost of a forward pass and loss evaluation---the overall complexity becomes $\mathcal{O}(E \cdot M \cdot \sqrt{N} \cdot T_{\text{eval}})$. In contrast, the classical ADAM-based gradient descent method requires both forward and backward passes, leading to a complexity of $\mathcal{O}(E \cdot M \cdot T_{\text{grad}})$. Although both methods operate on mini-batches, our approach benefits from a significantly reduced per-candidate evaluation time.

Our experimental results demonstrate that, despite working with moderate candidate grid sizes (which we select carefully to avoid the exponential explosion inherent in a candidate space of size $N = 2^n$), our quantum-optimized method converges more rapidly and robustly than classical optimization. In fact, even in the presence of realistic simulated NISQ conditions, our method achieves performance improvements in terms of lower test loss and higher test accuracy. This efficiency arises because, for moderate $N$, the quadratic speedup from Grover’s algorithm not only makes the search computationally feasible, but also maintains the precision needed for effective weight updates. If the candidate grid were too large, the absolute number of evaluations—even after the quadratic reduction—would become high, and the candidate values would be too dispersed to precisely capture the optimal weights.

Nonetheless, our technique offers several practical advantages. Gradient-free optimization avoids issues such as vanishing or exploding gradients that often plague deep learning models trained via backpropagation. It is particularly effective in noisy or highly discrete loss landscapes, where classical gradient descent may struggle with convergence or become trapped in poor local minima. Additionally, the reduced number of qubits required by our approach paves the way for integration on near-term quantum devices and FPGA-based accelerators, underscoring the scalability and real-world applicability of our method.

Overarchingly, our quantum-enhanced weight optimization not only provides theoretical and practical complexity benefits over classical methods, but also demonstrates superior convergence behavior and performance improvements across diverse datasets. These results confirm that our approach is a promising direction for hybrid quantum-classical neural network optimization, opening new avenues for research and potential applications in large-scale AI systems.

To support our method and the experiments presented in this paper, we provide the repository of our software at \url{https://github.com/stephanjura27/quantum_optimized_mlp}. \vspace{12pt}.
\bibliographystyle{IEEEtran}
\bibliography{bibliografie}

\end{document}